# Revivals and entanglement from initially entangled mixed states of a damped Jaynes – Cummings model

**R. W. Rendell and A. K. Rajagopal**
Naval Research Laboratory, Washington DC 20375-5320

**ABSTRACT**

An exact density matrix of a phase-damped Jaynes-Cummings model (JCM) with entangled Bell-like initial states formed from a model two-state atom and sets of adjacent photon number states of a single mode radiation field is presented. The entanglement of the initial states and the subsequent time evolution is assured by finding a positive lower bound on the concurrence of local 2x2 projections of the full 2x∞ JCM density matrix. It is found that the time evolution of the lower bound of the concurrence systematically captures the corresponding collapse and revival features in atomic inversion, relative entropies of atomic and radiation, mutual entropy, and quantum deficit. The atom and radiation subsystems exhibit alternating sets of collapses and revivals in a complementary fashion due to the initially mixed states of the atom and radiation employed here. This is in contrast with the result obtained when the initial state of the dissipationless system is a factored pure state of atom and radiation, where the atomic and radiation entropies are necessarily the same. The magnitudes of the entanglement lower bound and the atomic and radiation revivals become larger as both magnitude and phase of the Bell-like initial state contribution increases. The time evolution of the entropy difference of the total system and that of the radiation subsystem exhibits negative regions called "supercorrelated" states which do not appear in the atomic subsystem. Entangled initial states are found to enhance this supercorrelated feature. Finally, the effect of phase damping is to randomize both the subsystems for asymptotically long times.

## I. INTRODUCTION

The Jaynes - Cummings model (JCM) [1, 2, 3] is an exactly soluble quantum mechanical model of interaction between two dissimilar quantum systems - a single-mode radiation field and a two-level atom. There are many ongoing experimental and theoretical investigations including the relevant damping mechanisms (e.g. cavity losses, spontaneous emission) [4] providing precise tests of the JCM and its generalizations [5]. The different versions of the JCM form bases for a vast array of models for current experiments on the foundations of quantum mechanics involving entangled states, new ideas of quantum optics, and novel device structures such as the micromaser, single-atom laser, etc. [6-9]. This model has acquired further significance because it contains subtle features of quantum entanglement [10] that are interaction-strength-dependent.

The effect of different types of dissipation on the JCM has been studied both analytically and numerically [11-20]. The JCM and related models with dissipation have more recently attracted interest in studies of quantum computation [21] and quantum non-demolition measurements [22]. Among these, phase damping (which occurs when there is no energy exchange between the system and environment) is relevant to the decoherence of qubits, in the usual situation where decoherence takes place on a shorter time scale than for energy dissipation [23].

In all of these investigations, the initial system density matrix is taken to be a product of pure states of the factored form, $\rho_F(0) = \rho_A(0) \otimes \rho_R(0)$. The atom is often taken to be in the excited pure state $\rho_A(0) = |2\rangle\langle 2|$ and the laser radiation is taken to be a pure state density matrix of a coherent state: $\rho_R(0) = |\alpha\rangle\langle\alpha|$ with $|\alpha|^2 = N,$ the mean number of photons. This will evolve in time under the action of JCM dynamics and in general it will not remain in the factored form. Quantum entanglement of a composite system is defined by the inability to express its density matrix in terms of a sum of the density matrices of the components (A, B) of the separable form $\sum_i w_i \rho_i(A) \otimes \sigma_i(B),$ where the $w_i$'s are positive fractions whose sum is unity. The factored initial state is a special case of this more general form. The subsequent time evolution of the factored $\rho_F(0)$ given above is in general not manifestly separable.

The purpose of this paper is to consider initially entangled mixed states of photons and the atom for which an exact solution for a phase-damped JCM is found. Thus the present work sheds light on the effects of entangled initially mixed states of the coupled system and how this is affected by phase damping. The initial state used in our work is manifestly not factored into atomic and radiation terms, but more significantly we demonstrate further that it is not separable in the more general sense stated above. JCM is a $2 \times \infty$ system and there is presently no known method for computing its entanglement. The entanglement of formation or, equivalently, the concurrence [24] can quantify the entanglement of 2x2 systems as well as other particular finite systems with special symmetries. However we calculate a lower bound for the concurrence of the JCM system in terms of an average over all possible 2x2 projections whose concurrence can be calculated. When this lower bound is found to be positive for the parameters defining the initial state, the system is assured to be not only not factored but in fact truly entangled. This method is also used for examining the entanglement at subsequent times. The effects of the initial entangled state on the subsequent time evolution on the atomic inversion and various entropic functionals are then studied in an effort to discern which system property best captures the entanglement features. It is important to point out at the outset that for the initially factored pure states of the atom and radiation in the discussion found in the literature, the entropies of the atom and radiation subsystems are identical (see [7], for example). The results obtained in the present work using initially factored and entangled mixed states will be shown to differ from this as well as in other ways.

By employing the Poisson sum formula and the stationary phase method, the collapse and revivals of the atomic inversion in the original JCM were explained [25]. The results of similar calculations are used here to understand the dependence of these properties on the initial entangled mixed states. Correlations in the combined atom – radiation system is studied by examining the von Neumann entropies of the atom and radiation subsystems. Comparison of these with the concurrence lower bound (CLB) reveals the role of entanglement in the structures of the various trends found in these quantities. This paper is thus a contribution to the understanding of the consequences of preparation of the initial state of the dissipative JCM and the entanglement characteristics of two disparate systems.

In sec. II, we describe the phase damped dissipative JCM and also specify the initial condition on the density matrix and the lower bound of its entanglement. In sec. III, we give the exact solution of the density matrix equation with the entangled initial condition set up in sec. II. From this we deduce expressions for the marginal density matrices of the atom and the radiation field. We analyze the atomic inversion properties using the methods in ref.[25] which clearly show the dependence on the nature of the initial state as well as the entanglement features as exposed in the time-dependence of the CLB. In sec. IV, the expressions for the von Neumann entropies are defined for the reduced density matrices of the atom and the radiation. Also in sec. IV, the correlation properties are studied by examining the "quantum deficit" [26], defined as $D(A,R) = S_d(A,R) - S(A,R)$, where $S_d(A,R)$ is the von Neumann entropy of the decohered density matrix and $S(A,R)$ is that of the exact composite density matrix. Other indicators, the entropy differences of S(A,R) and S(A), S(R) and S(A) + S(R) - S(A,R) where S(A) is the entropy of the marginal density matrices of the atom and S(R), that of the radiation, are also examined and compared with the corresponding CLB. In Section V, we discuss the results obtained by means of graphical presentations of the formal expressions obtained in sec. IV for particular values of the parameters of the model. In the final section VI, a summary of the results and conclusions are given. In the Appendix, we present an outline of the calculation of the CLB to examine the time evolution of the entanglement using the exact solution obtained here.

## II. FORMULATION OF THE PROBLEM

The Hamiltonian of the JCM concerns a two-level system (TLS) interacting with a single mode of quantized radiation (boson) field of frequency, $\omega$, described in terms of the usual creation, $\hat{a}^\dagger$, and destruction, $\hat{a}$, operators of the photon field. The TLS is represented by the z-component of the Pauli matrix operator with the energy separation of the two atomic levels $\omega_o$, and their mutual interaction is expressed in the rotating wave approximation:

$$H = \hbar\omega \hat{a}^\dagger \hat{a} + \frac{\hbar\omega_o}{2}\hat{\sigma}_z + \hbar\kappa(\hat{a}^\dagger \hat{\sigma}_- + \hat{a}\hat{\sigma}_+). \qquad (1)$$

Here $\kappa$ is the dipole-interaction strength between the TLS and the field. Exact solutions of this interacting system are known [1 - 3] and their density matrices are also known [10] when the initial state is factored. Here exact analytic expressions are given for an entangled initial mixed state in the presence of phase damping for the resonant case where $\Delta\omega = \omega - \omega_0 = 0$. A non-unitary time-evolution of an initially prescribed density matrix of the system, $\rho(0)$, describing a special type of dissipation is given by the equation

$$\frac{d}{dt}\rho = \frac{1}{i\hbar}[H,\rho] + \frac{\gamma}{2}(\hat{\sigma}_z \rho \hat{\sigma}_z - \rho). \qquad (2)$$

The first term is the unitary evolution governed by the JCM Hamiltonian, eq.(1), and the second term is the dissipative, non-unitary part, with $\gamma$, a positive parameter, representing a phase decoherence of the atomic system induced by the environment. Phase damping occurs due to dephasing interactions that might arise, for example, from elastic collisions in an atomic vapor. Equation (2) is in the Lindblad form [27, 28] and therefore preserves all the required properties of the density matrix - hermiticity, trace-class, and positive-semi-definiteness. It will be solved here exactly in an analytic form with an entangled initial mixed state in parallel to the exact solution with a factored initial state as in [1, 2, 3, 10].

The normalized initial state density matrix is constructed by supplementing the usual factored initial state with a second "Bell" piece, $\Delta\rho$, formed from a sum of Bell-like states involving the two atomic states (1 stands for the ground and 2 for the excited state of the atom) and two adjacent photon number states:

$$\rho(A,R;t=0) = (1-\lambda)\rho_A(0) \otimes \rho_R(0) + \lambda\Delta\rho(A,R;t=0). \tag{3a}$$

$$\Delta\rho(A,R;t=0) = \sum_{n=0}^{\infty} p(n)|\Psi_n(1,2)\rangle\langle\Psi_n(1,2)|, \tag{3b}$$

where $|\Psi_n(1,2)\rangle \equiv (|n,1\rangle + a|n+1,2\rangle)/\sqrt{(1+|a|^2)}$, and $a = |d|\exp(-i\varphi)$.

It will be seen that nonzero phase $\varphi$ is necessary in order for the "Bell" piece to modify the system dynamics. For this reason we henceforth call it the "Bell phase". Depending on the values of the model parameters, this system may be factorizable, separable, or entangled. The entanglement parameter, $\lambda$ $(0 \leq \lambda \leq 1)$, in eq.(3a) allows us to specify initial states ranging from completely factored $(\lambda = 0)$ to completely "Bell", eq.(3b)$(\lambda = 1)$. In contrast to earlier work (see for example, [7]) we employ here a mixed two-state atomic density matrix given by $\rho_A(0) = p_{11}|1\rangle\langle 1| + p_{22}|2\rangle\langle 2|$ and a mixed state radiation density matrix in photon number representation, $\rho_R(0) = \sum_{n=0}^{\infty} p(n)|n\rangle\langle n|$. We illustrate all our results using the laser source [10], with representative values of the mean number of photons, N, the photon number distribution being a Poisson distribution, $p(n) = (N)^n \exp(-N)/n!$. This distribution has a peak at $n = N$ and vanishes for large n. $p_{11}$ and $p_{22}$ are the components of the two-state density matrix of the atom with $p_{11} + p_{22} = 1$ and $\sum_{n=0}^{\infty} p(n) = 1$. The initial state, eq.(3a), however must be examined as to its entanglement status for reasons to be discussed presently. When $\lambda \neq 0$, it manifestly does not factor into an atomic and a radiation term; but the possibility remains that it may be separable in the more general form $\sum_i w_i \rho_i(A) \otimes \sigma_i(B)$, and therefore may or may not be entangled. There is at present no known method for computing the entanglement of such $2 \times \infty$ JCM systems. However following the method in [29] we can determine a lower bound for the initial entanglement by considering projections onto 2x2 subspaces spanned by the radiation states $|n\rangle$ and $|n+1\rangle$ in eq.(3). Since these projections involve only local actions, they cannot increase the entanglement on average [29]. The entanglement of each of these projections can be calculated using Wootter's formula for the concurrence of a 2x2 state [24] and the average over all values of n provides a lower bound for the system entanglement. If this lower bound is positive, then the system is entangled. Such a calculation places conditions on the parameters appearing in the initial state given by eq.(3) in order for it to be entangled. We also use the same method to examine the entanglement at finite times.

Although the actual entanglement for this state is not quantified, this calculation of a positive lower bound assures the system is truly entangled.

The initial states of the atom are coupled with the photons explicitly because the atom jumps from one state to the other by absorbing and emitting a photon, as represented by $\lambda \Delta \rho(t=0)$. This may be thought to arise from the laser source represented by a coherent state, and hence the interaction of the atom with the light source depends on the intensity of the source via its amplitude, as well as possible phase fluctuations. The parameters in this model are so chosen as to maintain the positive semi-definiteness of the density matrix. It should be emphasized that the preparation of the initial state as in eq. (3) is nontrivial both conceptually and experimentally. The purpose of the present work is to examine the physical consequences of such choices of entangled initial states.

## III. EXACT SOLUTION FOR THE DISSIPATIVE JCM DENSITY MATRIX

Introduce dimensionless time, interaction, and damping respectively: $\tau = \omega t$, $\kappa = \bar{\kappa}\omega$, and $\gamma = \bar{\gamma}\omega$. Then the solution to eq. (2) is sought in terms of the complete set of the photon number states $\{|n\rangle\}$, $n = 0, 1, \cdots$ and the atomic ground and excited states respectively given by $\{|i\rangle\}, i = 1, 2$: $\rho = \sum_{n,m=0}^{\infty} \sum_{i,j=1}^{2} |n\rangle|i\rangle(\langle n|\langle i|\rho|j\rangle|m\rangle)\langle j|\langle m|$. The equations for the various components are written down in a straightforward way. The reduced density matrices are obtained by taking partial traces: $\rho_R = Tr_A \rho(A,R)$ for the radiation and $\rho_A = Tr_R \rho(A,R)$ for the atom. These equations are solved by a Laplace transform method. With the initial conditions given by eqs. (3), the exact expressions for the components of the density matrix are given below.

$$\langle n|\rho_{11}(\tau)|n\rangle = \frac{1}{2}\langle n|\rho_{11}(0)|n\rangle\{1 + e^{-\bar{\gamma}\tau/2} W_+(n+1;\tau)\}$$

$$+ \frac{1}{2}\langle n+1|\rho_{22}(0)|n+1\rangle\{1 - e^{-\bar{\gamma}\tau/2} W_+(n+1;\tau)\} \quad (4)$$

$$+ 2\bar{\kappa}\sqrt{(n+1)}|\langle n|\rho_{12}(0)|n+1\rangle|Sin\varphi\, e^{-\bar{\gamma}\tau/2} V(n+1;\tau)$$

$$\equiv A(n,n;\tau).$$

$$\langle n+1|\rho_{22}(\tau)|n+1\rangle = \frac{1}{2}\langle n+1|\rho_{22}(0)|n+1\rangle\{1 + e^{-\bar{\gamma}\tau/2} W_+(n+1;\tau)\}$$

$$+ \frac{1}{2}\langle n|\rho_{11}(0)|n\rangle\{1 - e^{-\bar{\gamma}\tau/2} W_+(n+1;\tau)\} \quad (5)$$

$$- 2\bar{\kappa}\sqrt{(n+1)}|\langle n|\rho_{12}(0)|n+1\rangle|Sin\varphi\, e^{-\bar{\gamma}\tau/2} V(n+1;\tau)$$

$$\equiv B(n+1,n+1;\tau).$$

$$\langle n|\rho_{12}(\tau)|n+1\rangle = e^{-\bar{\gamma}\tau}\langle n|\rho_{12}(0)|n+1\rangle$$

$$+ i\sin\varphi|\langle n|\rho_{12}(0)|n+1\rangle|e^{-\bar{\gamma}\tau/2}\left(e^{-\bar{\gamma}\tau/2} - W_-(n+1;\tau)\right) \quad (6)$$

$$- i\bar{\kappa}\sqrt{n+1}(\langle n+1|\rho_{22}(0)|n+1\rangle - \langle n|\rho_{11}(0)|n\rangle)e^{-\bar{\gamma}\tau/2} V(n+1;\tau)$$

$$\equiv C(n,n+1;\tau).$$

$$\langle n+1|\rho_{21}(\tau)|n\rangle = \langle n|\rho_{12}(\tau)|n+1\rangle^*. \tag{7}$$

And,
$$\langle 0|\rho_{22}(\tau)|0\rangle = p(0)(1-\lambda)p_{22} \equiv B(0,0;\tau). \tag{8}$$

In the above, the initial density matrix elements are given from eq.(3) as
$$\langle n|\rho_{11}(0)|n\rangle = [(1-\lambda)p_{11} + \lambda q_{11}]p(n), \quad \langle n|\rho_{12}(0)|n+1\rangle = p(n)\lambda e^{-i\varphi}\sqrt{q_{11}q_{22}},$$
$$\langle n+1|\rho_{22}(0)|n+1\rangle = [p(n+1)(1-\lambda)p_{22} + p(n)\lambda q_{22}]. \tag{9}$$

Here the Bell-state parameters are $q_{11} = (1+|a|^2)^{-1}$, $q_{22} = 1 - q_{11}$. Also,
$$W_{\pm}(n+1;\tau) \equiv \left(Cos\tau E(n+1) \pm \frac{\bar{\gamma}}{2}V(n+1;\tau)\right);$$
$$V(n+1;\tau) \equiv \frac{Sin\tau E(n+1)}{E(n+1)}. \tag{10a}$$
$$E(n+1) = \sqrt{4\bar{\kappa}^2(n+1) - (\bar{\gamma}/2)^2} \tag{10b}$$

with the physical assumption that $4\bar{\kappa}^2 - (\bar{\gamma}/2)^2 > 0$.

For asymptotically large times, when $\bar{\gamma} \neq 0$ the components of density matrix are, as deduced from eqs. (4 - 9):
$$\langle n|\rho_{11}(\infty)|n\rangle = A(n,n;\tau = \infty)$$
$$= \frac{1}{2}(\langle n|\rho_{11}(0)|n\rangle + \langle n+1|\rho_{22}(0)|n+1\rangle)$$
$$= \langle n+1|\rho_{22}(\infty)|n+1\rangle = B(n+1,n+1;\tau = \infty), \tag{11}$$
$$\langle 0|\rho_{22}(\infty)|0\rangle = (1-\lambda)p_{22}p(0) = B(0,0;\tau = \infty), \quad \text{and}$$
$$\langle n|\rho_{12}(\infty)|n+1\rangle = 0 = C(n,n+1;\tau = \infty).$$

These expressions already reveal the dependence of the results on the initial state and will be used to obtain asymptotic behaviors of various quantities in the later discussions. For example, the initial atomic state evolves into a mixed state for asymptotically long times when there is dissipation, showing that dissipation randomizes the initial atomic state due to interaction with radiation. The last expression in eq. (11) in comparison with eq.(6) also shows that the system decoheres for long times on a time scale of the order of $2\bar{\gamma}^{-1}$.

The complete solution of the density matrix is a sum of 2x2 block diagonal forms in the representation, $\{|n,1\rangle, |n+1,2\rangle\}$, $n = 0,1,2,\cdots$:
$$\rho(A,R;\tau) =$$
$$\sum_n \begin{Bmatrix} A(n,n;\tau)|n,1\rangle\langle n,1| + B(n,n;\tau)|n,2\rangle\langle n,2| + \\ +C(n,n+1;\tau)|n,1\rangle\langle n+1,2| + C^*(n,n+1;\tau)|n+1,2\rangle\langle n,1| \end{Bmatrix} \tag{12}$$

This can be put into a diagonal representation in terms of entangled atom and photon number states as in the original JCM [1, 2, 3, 10] defined by:
$$\rho(\tau) = B(0,0;\tau)|0,2\rangle\langle 0,2| +$$
$$\sum_{n=0} \{|\varphi(n,a)\rangle\lambda(n,a;\tau)\langle\varphi(n,a)| + |\varphi(n,b)\rangle\lambda(n,b;\tau)\langle\varphi(n,b)|\}, \tag{13}$$

where
$$\begin{Bmatrix} \lambda(n,a;\tau) \\ \lambda(n,b;\tau) \end{Bmatrix} = \left( \frac{A(n,n;\tau) + B(n+1,n+1;\tau)}{2} \right) \pm D(n,n+1;\tau),$$

$$D(n,n+1;\tau) = \frac{1}{2}\left\{ (A(n,n;\tau) - B(n+1,n+1;\tau))^2 + 4|C(n,n+1;\tau)|^2 \right\}^{1/2}.$$

(14)

where we have introduced the following relationships:
$$|n,1\rangle = Cos\theta_n |\varphi(n,a)\rangle + e^{i\psi_n} Sin\theta_n |\varphi(n,b)\rangle,$$

$$|n+1,2\rangle = -e^{-i\psi_n} Sin\theta_n |\varphi(n,a)\rangle + Cos\theta_n |\varphi(n,b)\rangle,$$

where $\psi_n = \tan^{-1}[\operatorname{Im} C(n,n+1)/\operatorname{Re} C(n,n+1)]$

and $\tan 2\theta_n = -2|C(n,n+1)|/(A(n,n) - B(n+1,n+1)).$

(15)

Note $\{|0,2\rangle, |\varphi(n,i)\rangle\}, i = a, b$ and $n = 0,1,2,\cdots$ form a complete, orthonormal set of this JCM at each instant of time. It should be noted that the states given in eq.(15) are eigenstates of the density matrix which reduce to the eigenstates of the JCM Hamiltonian in the absence of dissipation [10]. It is important to observe that in this representation there is no decoherence, in contrast to the representation in eq.(12). From eq.(13), we obtain the eigenvalues of the initial state density matrix which must be non-negative. This condition places a constraint on the parameters if $\lambda \neq 0,1$ then $\lambda[(n+1)p_{11}q_{22} + Np_{22}(q_{11} - p_{11})] + Np_{11}p_{22} \geq 0$. For $\lambda = 0$ or $1$, there is no constraint.

From eqs.(4) and (5), the reduced density matrices of the radiation and the atom in this interacting system are:
$$\rho_R(\tau) = |0\rangle\{B(0,0;\tau) + A(0,0;\tau)\}\langle 0|$$
$$+ \sum_{n=0}^{\infty} |n+1\rangle P(n+1;\tau)\langle n+1|, \qquad (16)$$

where $P(n+1;\tau) = A(n+1,;n+1;\tau) + B(n+1,n+1;\tau).$

and $\rho_A(\tau) = |1\rangle w(1;\tau)\langle 1| + |2\rangle w(2;\tau)\langle 2|,$ (17)

where $w(1;\tau) = \sum_{n=0}^{\infty} A(n,n;\tau),$ and $w(2;\tau) = \sum_{n=0}^{\infty} B(n,n;\tau).$ Note that these reduced density matrices are properly normalized to unity, as they must be. From these we have the initial state probabilities of photon and atom states given by:

$$P(n;\tau = 0) = p(n)[(1-\lambda) + \lambda q_{11}] + p(n-1)\lambda q_{22}, \qquad (16a)$$
$$w(1;\tau = 0) = (1-\lambda)p_{11} + \lambda q_{11} = 1 - w(2;\tau = 0). \qquad (17a)$$

Since these density matrices are already diagonal in the respective representations, appropriate choice of the parameters maintains these eigenvalues to be positive and less than unity and thus bona fide renormalized probabilities. Unlike in the finite bipartite systems, the two marginal density matrices have very different physical features, a point that becomes more evident in subsequent sections.

In the next section, we will construct several functionals of the system density matrix which will be useful to compare the time evolution from entangled initial mixed states to that from factored initial mixed states, as displayed by the corresponding CLB.

## IV. FUNCTIONALS TO PROBE ENTANGLED INITIAL MIXED STATES

The properties of the exact time-dependent evolution from entangled initial mixed states can be displayed using several different functionals of the system density matrix including the CLB. These functionals emphasize different features of the system such as revivals, decoherence, mixedness, and entanglement. The exact solution is used to construct several of these, the most familiar for the JCM being the atomic inversion which displays the patterns of collapses and revivals. Numerical evaluations of several significant functionals of the density matrix are systematically displayed and analyzed in Sec. V.

First we calculate the CLB for both factored and entangled initial mixed states based on eq.(A4) of Appendix A. Figure 1a shows the CLB of the initial state as a function of the entanglement parameter $\lambda$, for various values of the probabilities of occupation of the atomic state for fixed values of the parameters of the "Bell" piece, Eq.(3b), of the initial density matrix, limited by the requirement of positive semi-definiteness of initial state density matrix. From this figure, we observe that for mixed initial atomic states with $p_{11} < 1$, the entanglement parameter shows a threshold for obtaining nonzero CLB. The CLB increases with $\lambda$ and becomes independent of the initial atomic state as we approach the "Bell" piece at $\lambda = 1$. In Fig.1b the same quantity is shown for different mean numbers of photons for fixed values of the parameters of initial states of atom and radiation. From this we find that the entanglement in the system is expected most significantly for low mean photon numbers over the entire range of the entanglement parameter. But for increasing N there is a progressively smaller region near the "Bell" piece where the CLB is positive. This is expected on physical grounds that for such large photon numbers, the system is behaving quite classically. In later numerical presentations of the results, we use N=5, 20 along with suitable values of $p_{11}, q_{11}, \bar{\kappa}, \bar{\gamma}, \varphi$ and $\lambda$. These mean numbers of photons are used to best illustrate the features depending on the property we wish to emphasize.

We employ the same method outlined in Appendix A to obtain the CLB in JCM as time evolves, which are displayed in Fig. 2a, b. In Fig.2a, the CLB is presented for systems with initially factored and entangled mixed states as a function of time for N=5. In an inset, the same cases with dissipation included $(\bar{\gamma} = 0.01)$ are also given. From this we observe that the factored initial state is less easily entangled than the entangled initial state for the system parameters chosen. The initially entangled states remain entangled in contrast to the initially factored state. In contrast, the initially factored state exhibits intermittent regions of zero lower bound, which may indicate separability of the composite states in these time regimes. In Fig.2b, the same quantities are calculated for N=20 which brings out these features in a more pronounced fashion. Later, we will relate the peaks and valleys in these figures to special features of atomic collapses and revivals as well as features associated with complementary behavior of the radiation subsystem.

Next we construct several different entropy functionals. The von Neumann entropy for the atomic subsystem is given by:

$$S_A(\tau) = -w(1;\tau)\ln w(1;\tau) - w(2;\tau)\ln w(2;\tau) \qquad (18a)$$

and that of the photon subsystem is

$$S_R(\tau) = -\sum_{n=0}^{\infty} P(n;\tau)\ln P(n;\tau). \qquad (18b)$$

When the initial state is factored with both the atom and radiation taken to be pure states as is often done in the literature [7], these two entropies must be the same as guaranteed by the Araki – Lieb theorem [30, see Phoenix and Knight]. This results in a single set of collapses and revivals. In the present case where the initial state is of the factored or entangled form with mixed initial states of atom and radiation, the atomic and radiation entropies are distinct and the corresponding collapses and revivals are found to alternate in a complementary way as will be shown presently.

The relative entropies $S(R|A) \equiv S(A,R) - S(R)$, $S(A|R) \equiv S(A,R) - S(A)$, and the mutual entropy $S(A:R) \equiv S(A) + S(R) - S(A,R) \equiv S(R:A) \geq 0$ are also displayed and compared with each other. In [26], it was shown that only when the conditional probabilities can be defined, the first two expressions are the traditional non-negative conditional entropies. Otherwise they do not have a definite sign. The mutual entropy as defined above is non-negative, and is zero if and only if the marginal states are not correlated. It is a measure of both quantum and classical correlation residing in the composite system. In classical context, these provide measures of correlation. In [26], the "quantum deficit", D(A, R) was defined as $D(A,R) = S_d(A,R) - S(A,R) \geq 0$. Here $S_d(A,R)$ is the von Neumann entropy of the decohered density matrix obtained from eq.(12) by dropping the off-diagonal terms C(n,n+1) and S(A,R) is the corresponding entropy of the full density matrix given by eq.(13). Both the full and decohered density matrices have identical atom and photon marginal density matrices. Thus D(A, R) is a measure of the quantum correlation in the system, because the decohered density matrix represents the classical part of the composite system. We will employ these here to discuss the entanglement properties of the JCM in conjunction with the CLB.

In order to provide accurate analytical expressions exhibiting the dependences on the parameters of the entangled initial states, we give a Poisson-sum analysis of the "atomic inversion" for the case with no dissipation:

$$\Delta w(\tau) = w(1;\tau) - w(2;\tau) = \sum_{n=0}^{\infty}\left(A(n,n;\tau) - B(n,n;\tau)\right) \qquad (20)$$

Following the Poisson summation procedure in [25], we express the final result in the following form:

$$\Delta w(\tau) = -\frac{1}{2}(1-\lambda)p_{22}\,p(n=0) + \Delta w_0(\tau) + \sum_{v=1}^{\infty}\Delta w_v(\tau) \qquad (21)$$

where

$$\Delta w_0(\tau) = \left\{ \begin{array}{l} \left[(1-\lambda)(p_{11}-p_{22})+\lambda(q_{11}-q_{22})\right]\cos 2\bar{\kappa}\tau\sqrt{N} \\ -(1-\lambda)(3p_{11}-p_{22})\bar{\kappa}\tau\dfrac{\sin 2\bar{\kappa}\tau\sqrt{N}}{2\sqrt{N}} \\ -\dfrac{3}{2}\lambda(q_{11}-q_{22})\bar{\kappa}\tau\dfrac{\sin 2\bar{\kappa}\tau\sqrt{N}}{2\sqrt{N}} \\ +2\lambda\sqrt{q_{11}q_{22}}\sin\varphi\sin 2\bar{\kappa}\tau\sqrt{N} \end{array} \right\} e^{-(\bar{\kappa}\tau)^2/2}$$

$$+ O\left(\dfrac{1}{N}\right)$$

(22)

gives the initial peak and collapse whereas $\Delta w_R(\tau) = \sum_{\nu=1}^{\infty} \Delta w_\nu(\tau)$, represents the successive revivals indexed by $\nu = 1, 2, \cdots$ which are well separated for sufficiently large $N$. Here $\Delta w_\nu(\tau)$ is given by

$$\Delta w_\nu(\tau) = \left(\dfrac{\bar{\kappa}\tau}{2\pi\sqrt{\nu^3}}\right)\left\{\dfrac{e^{-\frac{\bar{\kappa}^2}{2\pi^2\nu^2}(\tau-\tau_\nu)^2}}{\sqrt{\pi\overline{N}}}\right\}.$$

$$\left\{ \begin{array}{l} \left(\left[(1-\lambda)p_{11}+\lambda(q_{11}-q_{22})\right]\left(\dfrac{\tau}{\tau_\nu}\right)^2 - (1-\lambda)p_{22}\right)\cos\left(\dfrac{\bar{\kappa}^2\tau^2}{2\pi\nu}-\dfrac{\pi}{4}\right) \\ +2\lambda\sqrt{q_{11}q_{22}}\sin\varphi\left(\dfrac{\tau}{\tau_\nu}\right)^2\sin\left(\dfrac{\bar{\kappa}^2\tau^2}{2\pi\nu}-\dfrac{\pi}{4}\right) \end{array} \right\}$$

(23)

Here $\nu = 1, 2, \cdots$ and $\tau_\nu = (2\pi\nu\sqrt{N})/\bar{\kappa}$ are the familiar revival times. Since the Gaussian dominates at $\tau = \tau_\nu$, we obtain near this peak an expression in describing the spread around the revival. Note that there is another negligible term in these formulae, which comes from the stationary points with negative values of $\nu$ because the corresponding photon distribution is $\left\{e^{-\frac{\bar{\kappa}^2}{2\pi^2\nu^2}(\tau+\tau_\nu)^2}\Big/\sqrt{\pi\overline{N}}\right\}$. When $\lambda = 0$, $p_{11} = 0$ and $p_{22} = 1$, we recover the result in [25]. For general entangled mixed states, we can see the dependencies on atomic occupation $p_{11}$, Bell parameter $q_{11}$, and entanglement parameter $\lambda$, in eqs.(21 – 23). These expressions provide very good approximations to the exact results in eq.(20), but we do not plot them here for reasons of brevity. The expressions obtained above however give insight into understanding the collapses and revivals displayed in sec. V.

# V. DISCUSSION OF THE RESULTS

The results for particular values of the parameters are displayed in several figures. We discuss them in two broad groups in order to bring out the salient features of dissipation and initial state entanglement and their subsequent time dependence on the atomic and radiation properties as described by the JCM. In the first group, we examine the full composite system in various ways to understand the entanglement and correlation contained in the JCM. In the second group, we consider the properties of the atomic and radiation subsystems by examining their respective marginal density matrices.

In fig. 3a, we display the full system properties using mutual entropy, S(A:R), the quantum deficit, D, and the CLB, C, for the initially factored state (with atomic mixed state) to explore features of entanglement and correlation. Since D measures the difference between the entropies of decohered and the full system density matrix, it depicts the residual quantum correlation in the composite system. This figure reveals the collapse and revivals progressively in a manner similar to the CLB, C. It should be noted that the CLB vanishes or goes through a minimum in the regions where there are revivals exhibited in S(A:R) and D where they rapidly oscillate and the CLB goes through a maximum during each of the collapses.

Figure 3b shows the corresponding subsystem properties in terms of atomic inversion, atomic relative entropy, and radiation relative entropy. The purpose of this is to bring out unequivocally the physical features contained in each of these functionals and the information discerned from the mutual entropy and the quantum deficit, in relation to the CLB. The collapses and revivals of the atomic quantities $\Delta w$ and $S(A|R)$ are seen to occur at different times than the radiation quantity $S(R|A)$ and these alternate in a complementary way. Both of these sets of collapses and revivals are seen in the full system quantities C, D, and S(A:R). The collapses and revivals found in the atomic inversion $\Delta w$, correspond to the expressions obtained in eq.(22) for $v = 0,$ and eq.(23) for $v = 1, 2$ of the Poisson sum results. It was shown by Banacloche [30] that for the initially factored states where both the atomic and radiation states are pure, the collapses that appear in the atomic inversion are places where the atomic and radiation states become approximately factored again with no entanglement. By comparing the corresponding graphs in figs.3a and 3b, we see a similar phenomenon for the initially factored mixed case. Here this corresponds to the minima of the CLB, which are indicative of such partial disentanglements. However, with the initial mixed states used here, the entropies of the atomic and radiation subsystems are no longer identical and a separate set of radiation collapses and revivals alternate with the atomic ones.

Figures 4a,b are respectively the same type of information displayed for an initially entangled mixed states which are completely "Bell" $(\lambda = 1)$ in order to show the differences from the initially completely factored $(\lambda = 0)$ case shown in figs.3a,b. In this case, the values of the CLB are much larger than those for the initially factored state and also do not vanish as time evolves unlike its counterpart in fig.3a. Similar to the initially factored case, the CLB goes through a minimum at each of the atomic or radiation revivals. The values of C, D, and $S(A:R)$ are much larger than their counterparts in the initially factored case. We also display the dynamical effects of the Bell phase $\varphi$ on these quantities mentioned after eq.(3b) in sec.II. Explicitly, the density

matrix components of the JCM solution in eqs.(4, 5, 6) contain terms proportional to $\sin\varphi$, corresponding to atomic transitions simultaneously involving one-photon emission and absorption. The magnitudes of the revivals and the CLB increase to their respective maxima as the Bell phase increases towards $\pi/2$. The flat dashed lines correspond to $\varphi = 0$ where no such transitions occur. Note that D is always less than or equal to S(A:R) in figs.3a and 4a as shown in [26]. We conclude that the entanglement of atom and radiation in this model is reflected in the subsystems in complementary ways and both of these reside in D and S(A:R) in close conformity with C.

In contrast to classically correlated systems, it is possible in quantum entangled systems for the entropy of a subsystem to be larger than that of the total system. In the JCM, we find that this is possible for the radiation subsystem and exhibit this by the negative regions in the time evolution of $S(R|A)$ in figs.5a,b for the factored and entangled initial states. These negative regions are the special "supercorrelated" quantum features reported earlier in [10] for the initially factored state, which are forbidden in the classical case. The radiation relative entropy in figs.5a,b exhibit supercorrelated regions more frequently for the entangled initial state system than for the factored initial state system and this is further enhanced for smaller N. Furthermore each negative drop in the relative entropy is accompanied by an increase in C (shifted vertically in the figures for clarity). In fact, these states do not appear for the initially factored mixed state in contrast to the initially factored pure state. The supercorrelated regions have no counterpart in the atomic relative entropy for either pure or mixed initial states which in contrast have smaller subsystem entropy than that of the total system. Figures 3b,4b show the atomic relative entropies for mixed initial states. When dissipation is included, they ultimately approach asymptotic states, in conformity with eq.(11).

## VI. SUMMARY AND CONCLUDING REMARKS

In this paper we have examined the results of entangled initial mixed states and dissipation due to phase damping on the JCM in terms of the radiation-atom interaction strength, mean photon number of the radiation, along with the mixed initial state parameters. The initially entangled mixed state is constructed from a linear superposition of Bell-like states by combining the two states of the atom with sets of adjacent number states of the photon. The entanglement status of such an initial state is assured by finding a positive lower bound on the concurrence of local 2x2 projections of the full $2\text{x}\infty$ JCM density matrix. The same quantity is also computed to examine the entanglement properties of the subsequent evolution of the exact solution found for the JCM. These results are contained in two broad groups. One, properties of the full composite system are expressed in terms of the quantum deficit D, and the mutual entropy, S(A:R). Two, properties of the atomic and radiation subsystems in terms of the atomic inversion and the entropy differences, S(A,R) - S(A), and S(A,R) - S(R). They are all found to be correlated with the CLB to reveal the special features of this interacting many-body system. The atomic and radiation subsystems exhibit alternating collapses and revivals in a complementary fashion due to the initial mixed states of the atom and radiation employed here. The CLB tracks the features of the atom and radiation and their mutual correlations as displayed by mutual entropy, quantum deficit, and atomic inversion. The magnitudes of these entropic functions and the corresponding CLB increase as the entanglement parameter $\lambda$ increases and as the Bell-phase $\varphi$ increases from 0 to $\pi/2$. The initial state specificity persists over a time scale of $2\bar{\gamma}^{-1}$ in the dissipative cases whereas in the non-dissipative cases, they exhibit revivals. The revivals are present even with dissipation but with reduced amplitudes and both subsystems are randomized at asymptotically long times.

Another significant feature of the effect of initial state specification and dissipation is revealed in figs. 5a,b. Figure5a clearly shows negative regions as time progresses corresponding to radiation subsystem entropies which are greater than the total system entropy. This is in conformity with the earlier report using factored initial states [10], where it was called "supercorrelated". Initially entangled states greatly enhance this supercorrelated feature. The corresponding atomic subsystem relatiuve entropies in figs.3b,4b do not exhibit such supercorrelated states because the radiation subsystem is an infinite-state photon system in contrast to the two-state atomic subsystem. The corresponding dotted curves are obtained when a small amount of dissipation, is included which show how the "supercorrelated" region at small initial times survive but are quickly lost soon after. Moreover, the dissipation effect is to randomize the system such that for longer times the system is approaching the chaotic state of the atom as is also revealed in eqs.(9, 11).

The quantum entanglement in the JCM arises from interaction between the two disparate subsystems $(2 \times \infty)$ and therefore depends on the interaction strength between the two, in contrast to quantum entanglement of bipartite (2x2) or (2x3) systems [31, 32].

**Acknowledgments:** Besides thanking the Office of Naval research for partial support of this work, the authors thank Dr. Peter Reynolds of the same office for additional support.

# APPENDIX A : Lower bound on entanglement in JCM

We give the outline of a calculation of the lower bound on the entanglement for any time by computing the concurrence for a projection of the $2 \times \infty$ JCM solution onto an effective $2 \times 2$ system for each photon state n. The projection operator used is

$$\Pi_n \equiv (|1\rangle\langle 1| + |2\rangle\langle 2|) \otimes (|n\rangle\langle n| + |n+1\rangle\langle n+1|). \tag{A1}$$

This results in a 2-qubit-like density matrix appearing in Wootter's calculation [25] of the concurrence:

$$\rho_{\text{Pr}}(A, R; \tau) \equiv \frac{\Pi_n \rho(A, R; \tau) \Pi_n}{tr\{\Pi_n \rho(A, R; \tau) \Pi_n\}} = \frac{1}{T_n} \begin{pmatrix} v & 0 & 0 & 0 \\ 0 & \omega & z & 0 \\ 0 & z^* & x & 0 \\ 0 & 0 & 0 & y \end{pmatrix}_n \tag{A2}$$

where

$$v = B(n, n; \tau), \quad \omega = B(n+1, n+1; \tau), \quad x = A(n, n; \tau), \quad y = A(n+1, n+1; \tau),$$

$$z = C(n, n+1; \tau), \text{ and } T_n = (v + \omega + x + y)_n.$$

Here A, B, C are as given in the text by eqs.(4, 5, 6) respectively.
The concurrence lies between 0 and 1 and is given by the formula [25]

$$C(n; \tau) = \frac{2}{T_n} (|z| - \sqrt{vy})_n, \text{ if } \sqrt{\omega x} \geq |z|,$$

$$= \frac{2}{T_n} (\sqrt{\omega x} - \sqrt{vy})_n, \text{ if } \sqrt{\omega x} \leq |z|. \tag{A3}$$

This is then averaged over the probability distribution, $P_c(n)$, associated with finding such a projection:

$$C(\tau) = \sum_{n=1}^{\infty} C(n; \tau) P_c(n), \text{ where } P_c(n) = T_n \Big/ \sum_{n=1}^{\infty} T_n. \tag{A4}$$

Such mean values are displayed in figs. 1a,b for the initial state and figs. 2a,b for time evolution in order to obtain information about the entanglement status of the system revealed by the lower bounds on the concurrence given by eq.(A4).

# Figure Captions

Fig.1a The concurrence lower bound of the initial state as a function of $\lambda$ for N=2, $q_{11} = 0.5, \varphi = \pi/6, \bar{\kappa} = 1, \bar{\gamma} = 0$ and different $p_{11}$.

Fig. 1b The concurrence lower bound of the initial state as a function of $\lambda$ for $p_{11} = 1, q_{11} = 0.5, \varphi = \pi/6, \bar{\kappa} = 1, \bar{\gamma} = 0$ and different N=2, 3, 5, and 20.

Fig. 2a Time dependence of the concurrence lower bound for N=5 with parameters $p_{11} = 0.8, q_{11} = 0.5, \varphi = \pi/6, \bar{\kappa} = 1, \bar{\gamma} = 0$. Inset shows dissipation effects with $\bar{\gamma} = 0.01$.

Fig. 2b Time dependence of the concurrence lower bound for N=20 with parameters as in fig.2a. Inset shows dissipation effects with $\bar{\gamma} = 0.01$. Both of these figures are for $\lambda = 0 (\text{bottom}), 0.9 (\text{middle}), 1.0 (\text{top})$ displaying the dependence on the initial state entanglement.

Fig.3a Evolution from an initially factored state of composite system quantities concurrence lower bound, C, quantum deficit, D, and mutual entropy, S(A:R) with N=20, $\lambda = 0, p_{11} = 0.8, \bar{\kappa} = 1, \bar{\gamma} = 0$.

Fig.3b Evolution from an initially factored state of marginal system quantities atomic inversion, $\Delta w$, $S(A|R)$ and $S(R|A)$ for the same parameters as in fig.3a. Dotted curves are for $\bar{\gamma} = 0.05$.

Fig.4a Evolution from an initially entangled state of composite system quantities concurrence lower bound, C, quantum deficit, D, and mutual entropy, S(A:R) with N=20, $\lambda = 1, q_{11} = 0.5, \bar{\kappa} = 1, \bar{\gamma} = 0$ for values of the Bell-phase $\varphi = 0 \text{ (dashed)}, \pi/6 \text{ (solid)}, \pi/2 \text{ (dotted)}$.

Fig.4b Evolution from an initially entangled state of marginal quantities atomic inversion, $\Delta w$, $S(A|R)$ and $S(R|A)$ for the same parameters as in fig.4a. Dotted curves are for $\bar{\gamma} = 0.05$.

Fig. 5a Entropy difference $S(R|A) \equiv S(A,R) - S(R)$, showing supercorrelated states for both $\lambda = 0$, with $p_{11} = 0$ and 0.8, and $\lambda = 1$, with $q_{11} = 0.5, \varphi = \pi/6$ (independent of $p_{11}$). The other parameters used are N=20, $\bar{\kappa} = 1$, and $\bar{\gamma} = 0 (\text{solid}), 0.05 (\text{dotted})$.

Fig. 5b Entropy difference $S(R|A) \equiv S(A,R) - S(R)$, showing supercorrelated states for both $\lambda = 0$, with $p_{11} = 0$ and 0.8, and $\lambda = 1$ (independent of $p_{11}$). The other parameters used are the same as in fig.5a except N=5.

**Figure 1 (a)**

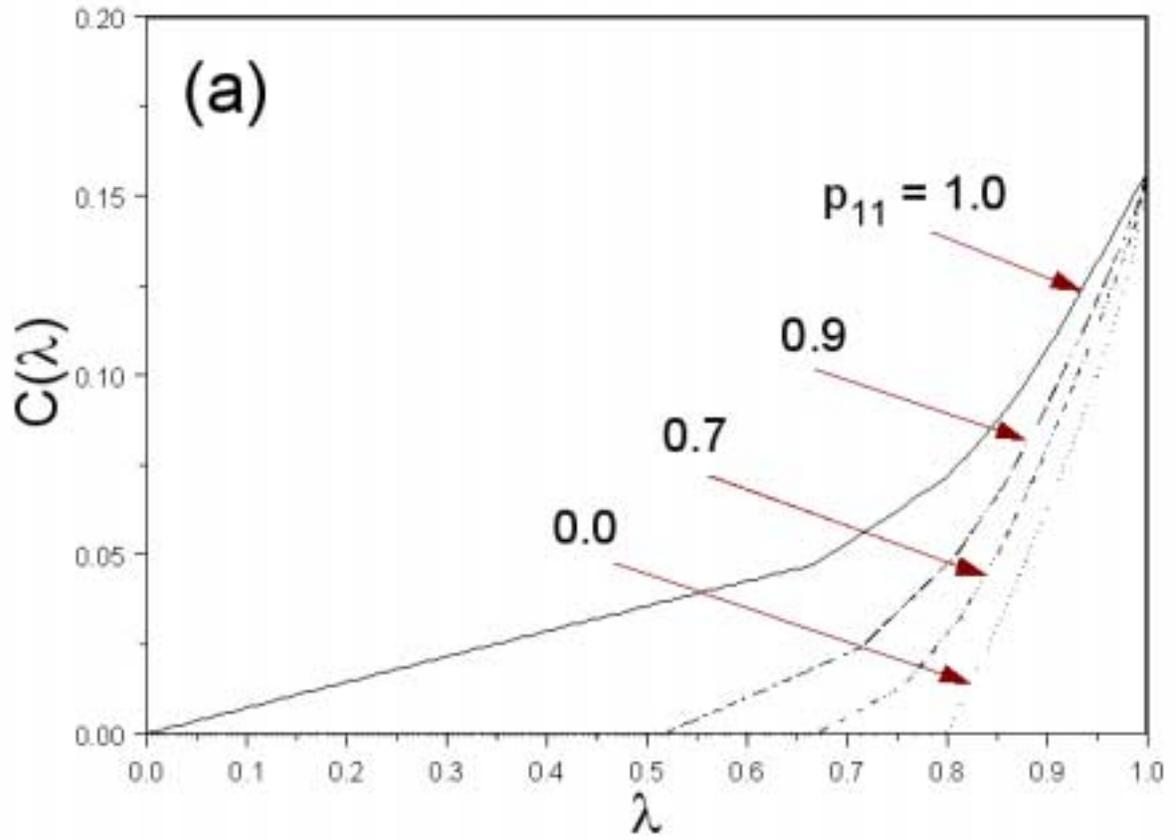

**Figure 1(b)**

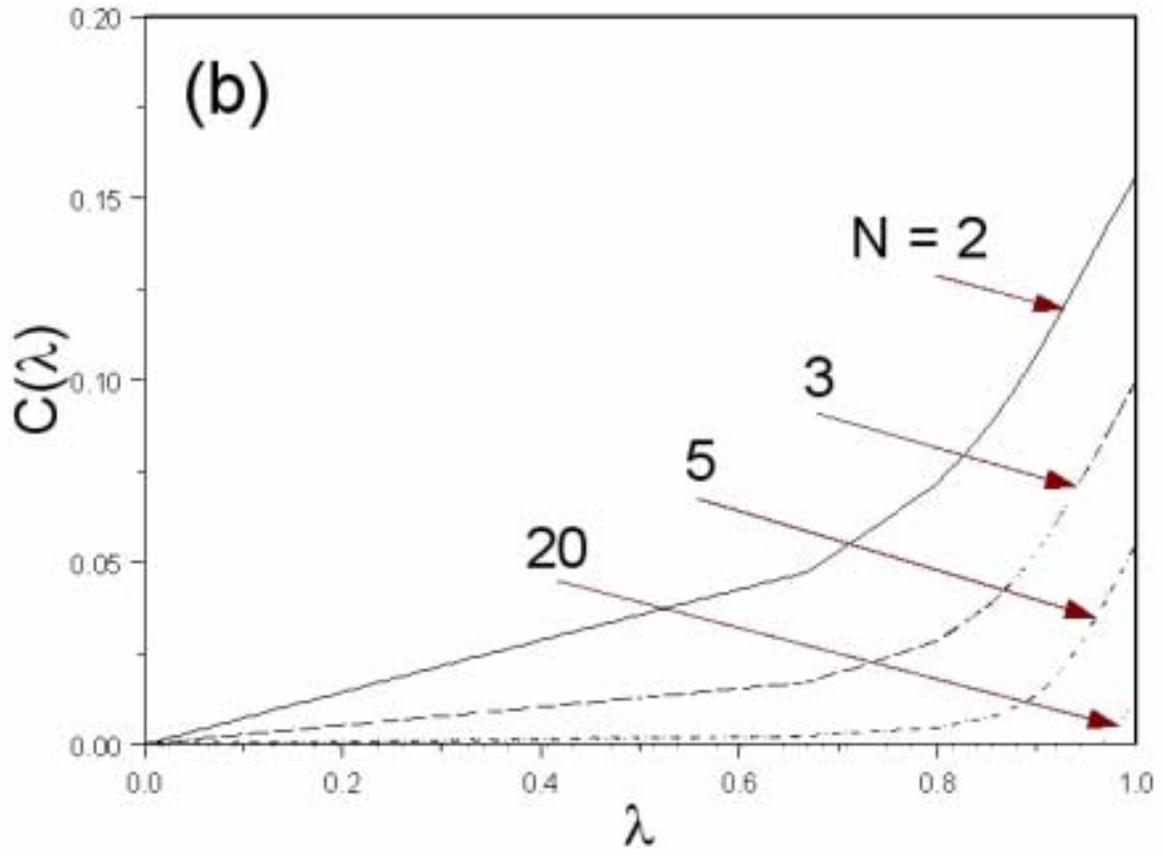

**Figure 2(a)**

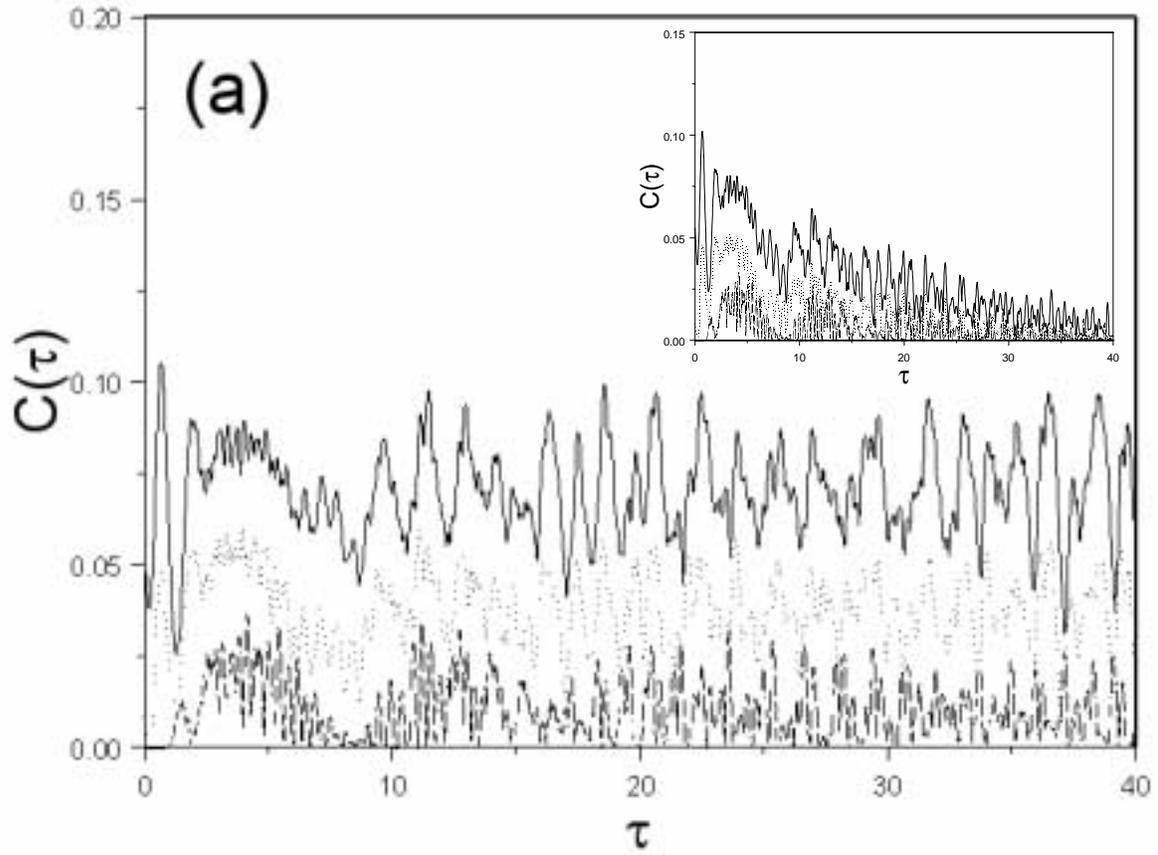

**Figure 2(b)**

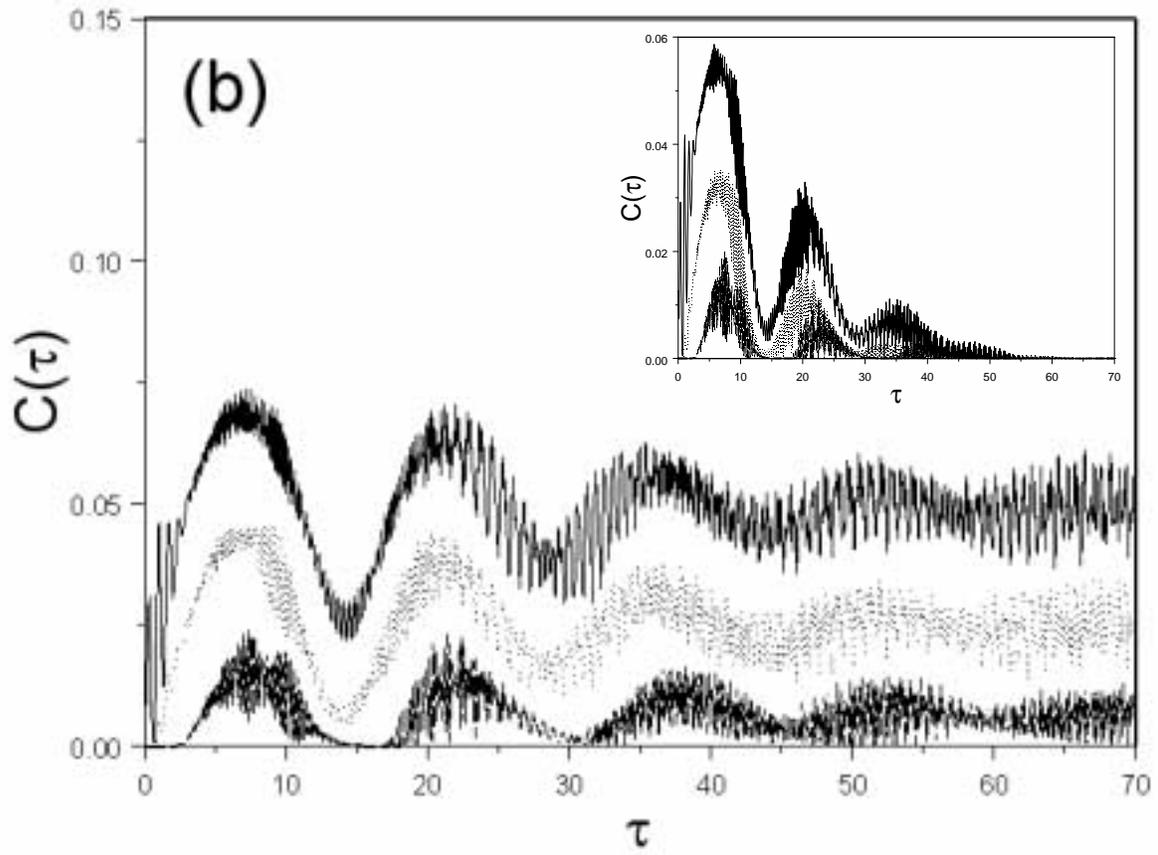

**Figure 3(a)**

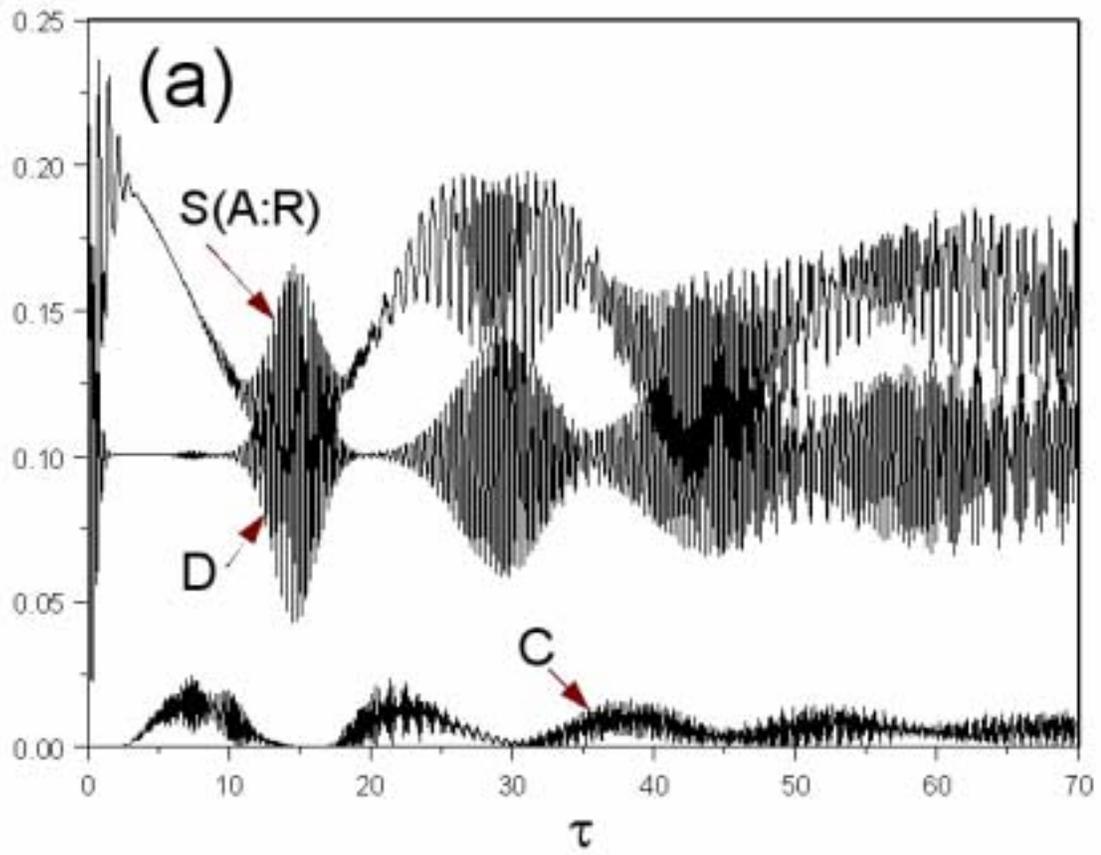

**Figure 3(b)**

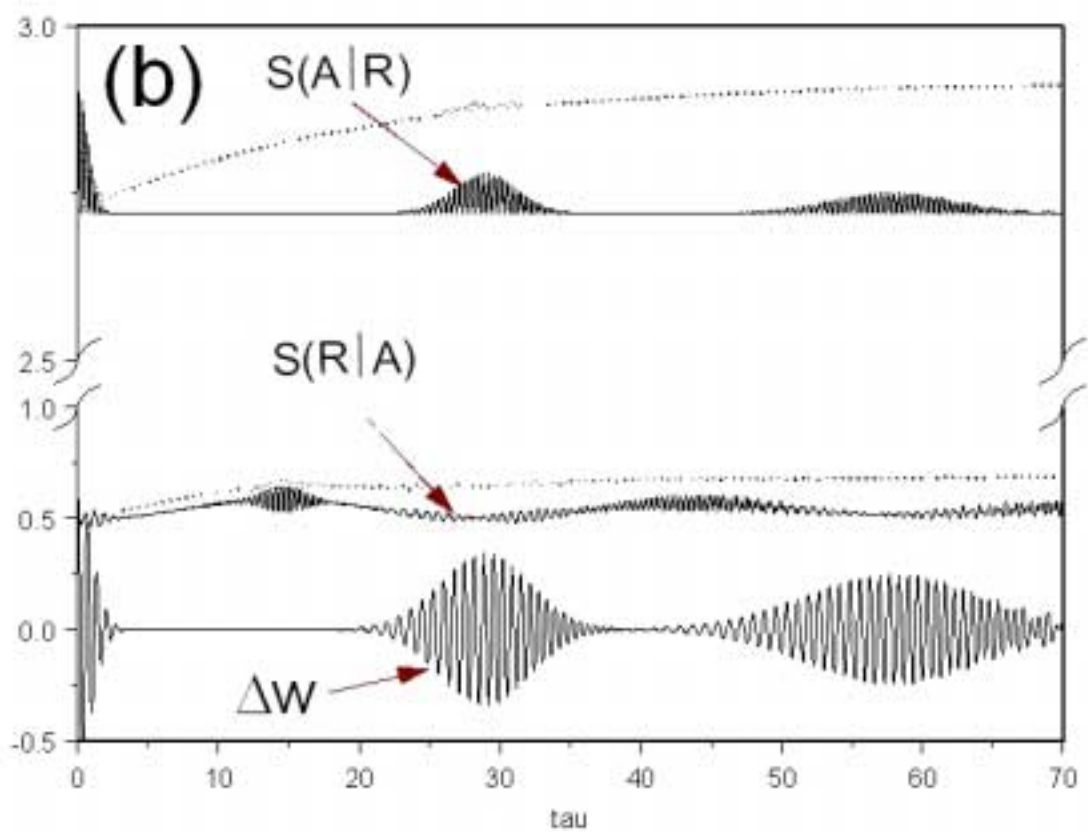

**Figure 4(a)**

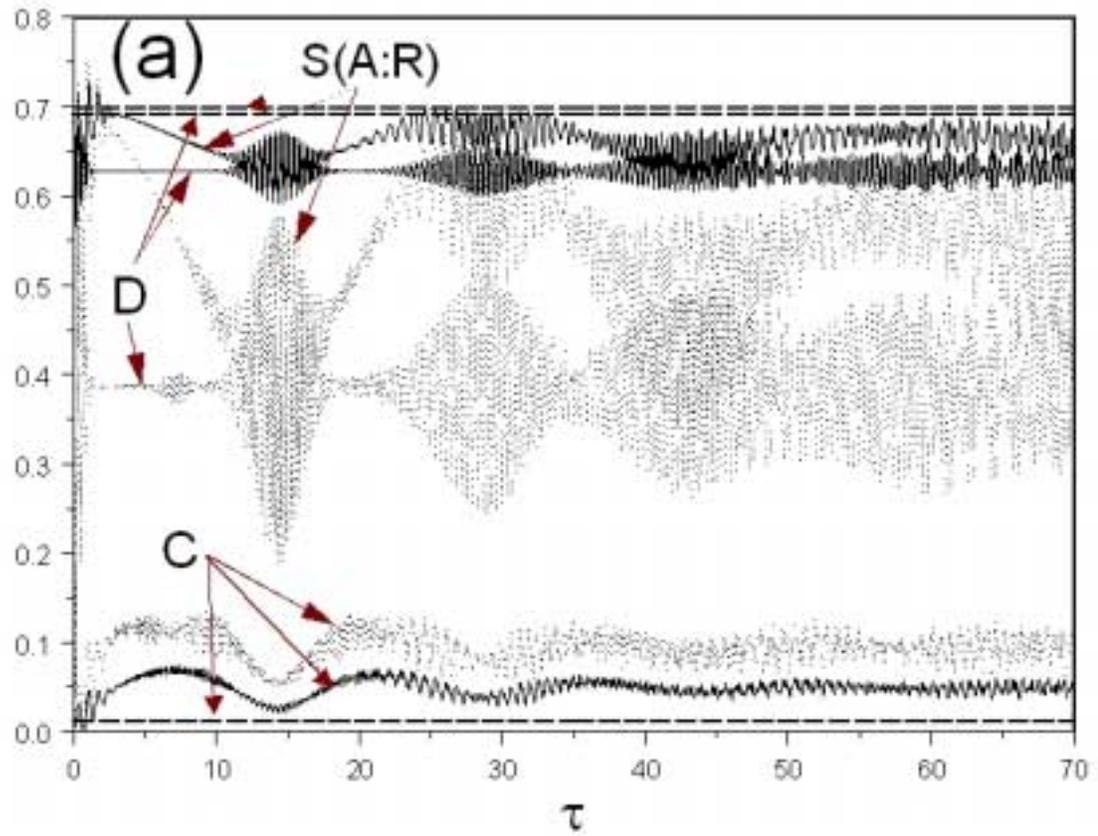

**Figure 4(b)**

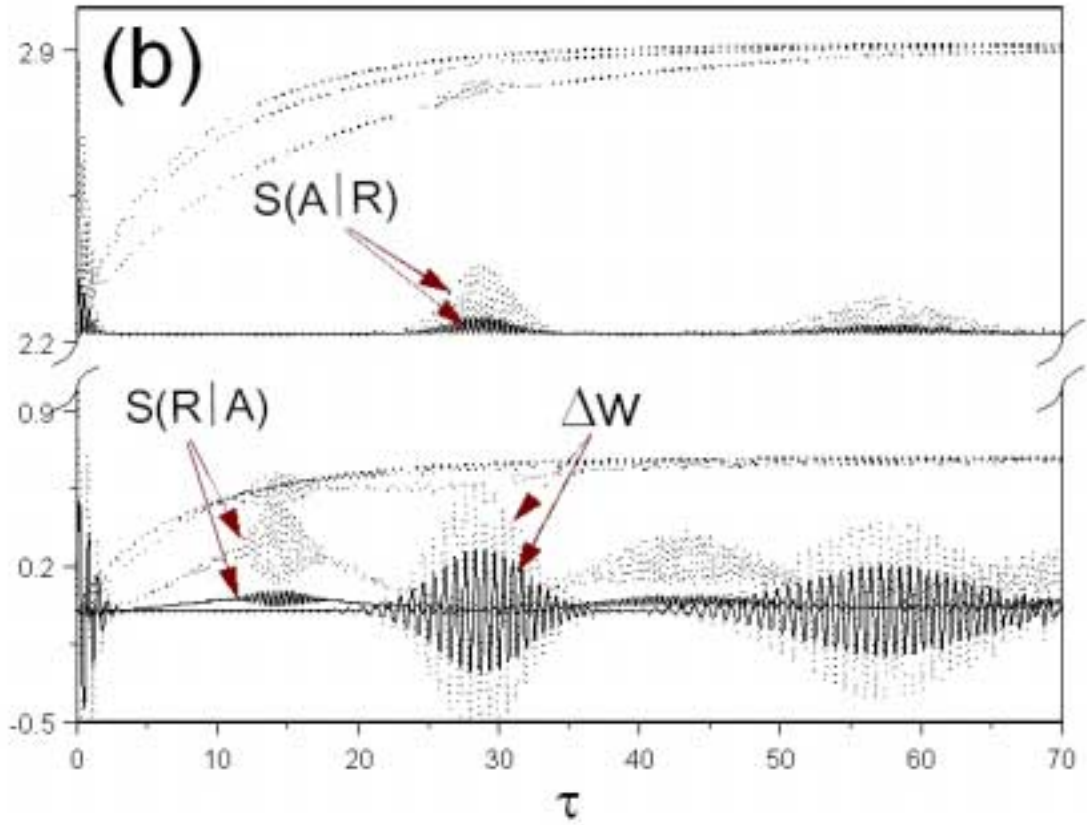

**Figure 5(a)**

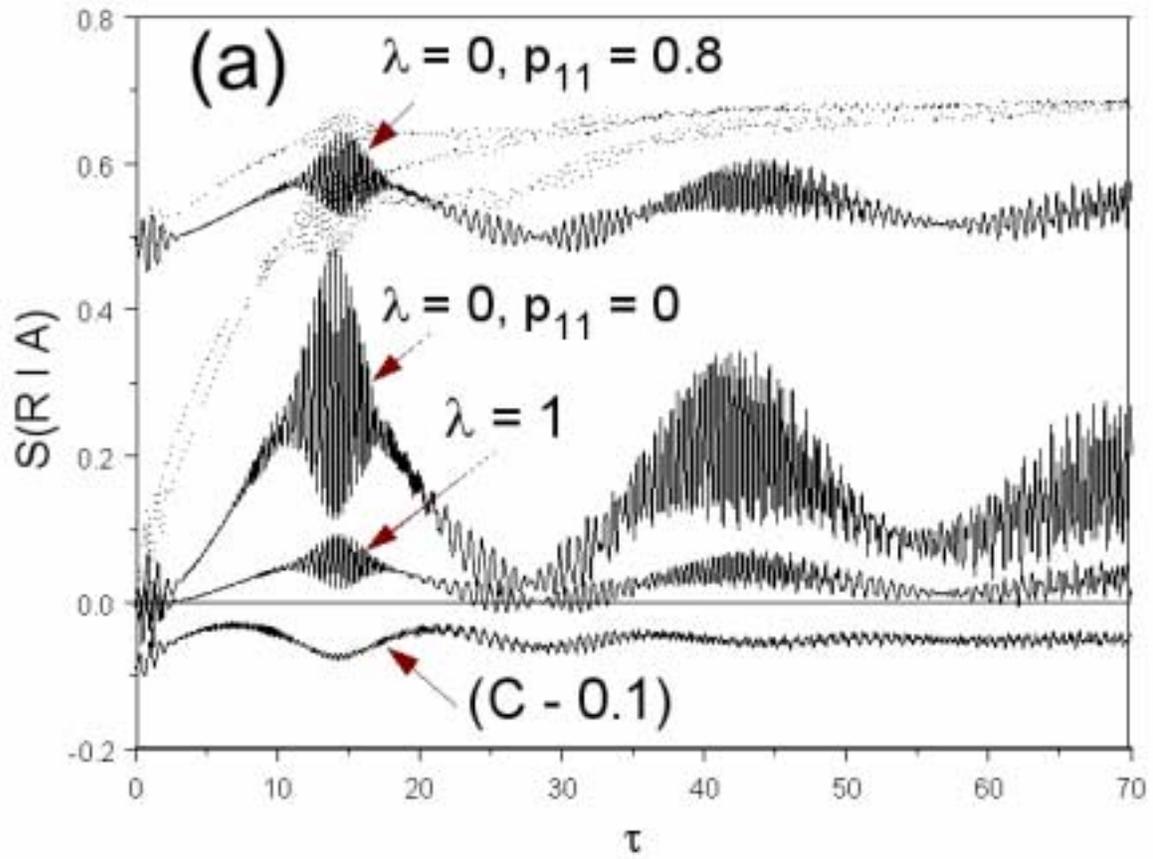

**Figure 5(b)**

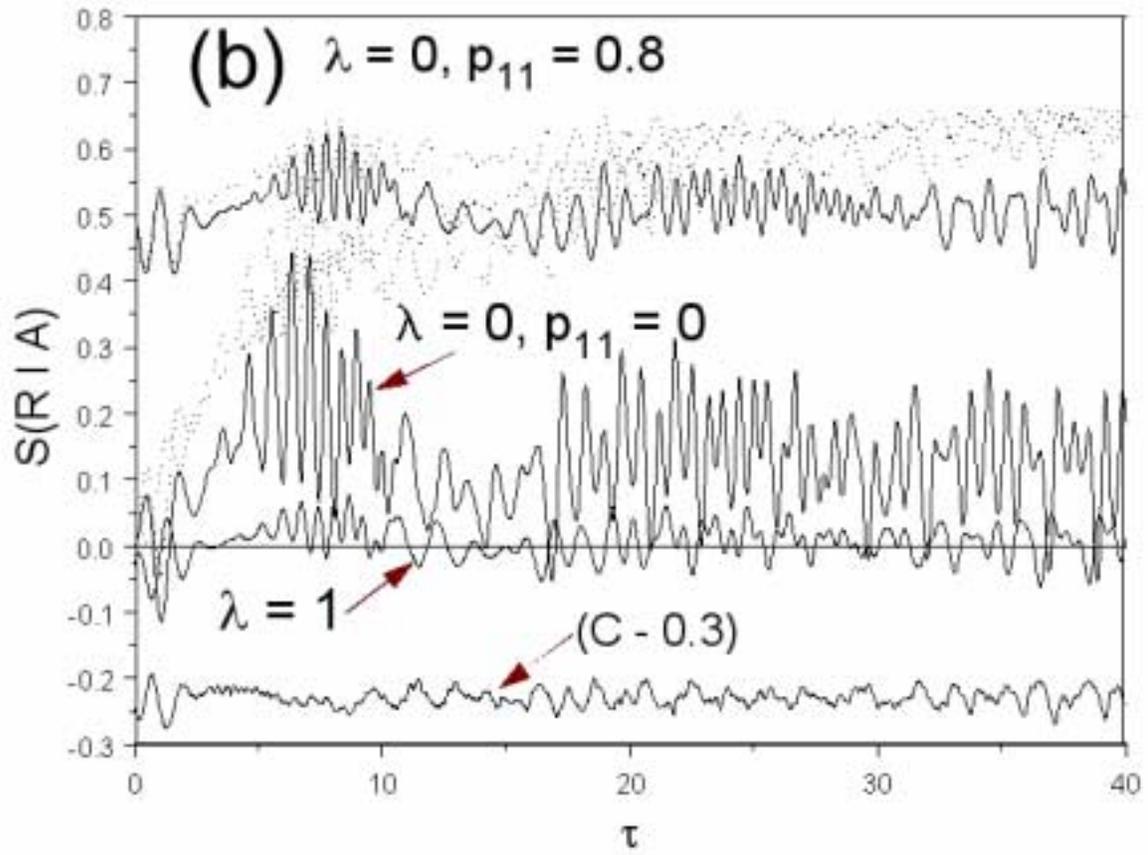